\documentclass{llncs}
\usepackage{url}
\usepackage{graphicx}
\usepackage[usenames,dvipsnames,svgnames,table]{xcolor}
\usepackage{hyperref}

\definecolor{linkblue}{RGB}{0,0,155}

\hypersetup{
   bookmarksnumbered,
   colorlinks={true},
   pdfstartview={FitH},
   citecolor={linkblue},
   linkcolor={linkblue},
   urlcolor={linkblue},
}

\begin{document}

\title{Reliable Granular References to\\Changing Linked Data}

\author{
Tobias Kuhn\inst{1}, Egon Willighagen\inst{2}, Chris Evelo\inst{2}, N\'uria Queralt-Rosinach\inst{3}, Emilio Centeno\inst{4}, Laura I. Furlong\inst{4}
}

\institute{
Department of Computer Science, Vrije Universiteit Amsterdam, Netherlands\\
\and
Department of Bioinformatics, NUTRIM, Maastricht University, Netherlands\\
\and
Department of Integrative Structural and Computational Biology, The Scripps Research Institute, La Jolla, USA\\
\and
Research Group on Integrative Biomedical Informatics (GRIB), Institut Hospital del Mar d'Investigacions Mèdiques (IMIM), Universitat Pompeu Fabra (UPF), Barcelona, Spain\\
}

\maketitle

\begin{abstract}
Nanopublications are a concept to represent Linked Data in a granular and provenance-aware manner, which has been successfully applied to a number of scientific datasets. We demonstrated in previous work how we can establish reliable and verifiable identifiers for nanopublications and sets thereof. Further adoption of these techniques, however, was probably hindered by the fact that nanopublications can lead to an explosion in the number of triples due to auxiliary information about the structure of each nanopublication and repetitive provenance and metadata. We demonstrate here that this significant overhead disappears once we take the version history of nanopublication datasets into account, calculate incremental updates, and allow users to deal with the specific subsets they need. We show that the total size and overhead of evolving scientific datasets is reduced, and typical subsets that researchers use for their analyses can be referenced and retrieved efficiently with optimized precision, persistence, and reliability.
\end{abstract}

\section{Introduction}

Datasets in general and Linked Data resources in particular play an increasingly important role in data-driven research, as exemplified by the datasets provided by WikiPathways \cite{Kutmon2016} and DisGeNET \cite{queralt2016disgenet}, and overarching initiatives such as Bio2RDF \cite{belleau2008bio2rdf}. Reproducibility and persistence have been ongoing concerns in this regard, as dataset identification and access has often been brittle and unreliable. Datasets based on Linked Data, as most types of datasets, are typically quite dynamic and change over time \cite{tzitzikas2008storage,fernandez2015diacron}, and capturing the data's provenance \cite{moreau2013provenance} is crucial for their proper interpretation and reuse. Moreover, as we will show, scientific data analyses typically use relatively small subsets of Linked Data resources, but we currently lack reliable methods to refer to such subsets.

In the context of the recent initiatives to promote FAIR data publishing \cite{wilkinson2016fair}, Linked Data can contribute to the requirement of interoperability across datasets. We argue that researchers should --- in papers as well as the software code for computational analyses --- be able to exactly specify what dataset they are using as input. Currently, the best researchers can do is to provide version numbers and bibliographic references in papers, like ``we used DisGeNET-RDF version 4.0 \cite{queralt2016disgenet}'', and to make the downloaded dataset explicit in the source code of their computational analyses, like in the following line of a Unix script:
\begin{quote}\scriptsize
\texttt{wget http://rdf.disgenet.org/download/v4.0.0/geneDiseaseAssociation.ttl.gz}\\
\texttt{\# Run analysis here}
\end{quote}
We can therefore identify the following two problems with the current practice of dataset references:
(1) Researchers can only specify at the dataset level which data they use as input. They cannot reliably point to the exact subset that is needed for a given analysis.
And (2) researchers cannot reliably refer to specific versions of evolving datasets; even with version numbers included, researchers cannot be sure that others can later retrieve exactly the same dataset to replicate the results.
We argue that we can address both problems with an approach of incremental dataset definitions based on the technologies of nanopublications and trusty URIs.

Nanopublications \cite{mons2011value} are tiny packages of Linked Data that come with provenance and metadata attached \cite{groth2010anatomy}. In previous work, we showed how identifers based on cryptographic hashes, called trusty URIs \cite{kuhn2014trusty,kuhn2015tkde}, can be used in combination with nanopublications to make them (and their entire reference trees) immutable and verifiable, two important properties for scientific data. In contrast to other proposals for data citations \cite{task2013out}, such a cryptography-empowered approach can provide us with strong technical --- rather than weaker organizational --- guarantees with respect to the integrity and original state of datasets.

\begin{figure}[tb]
\begin{center}
\includegraphics[width=0.9\textwidth]{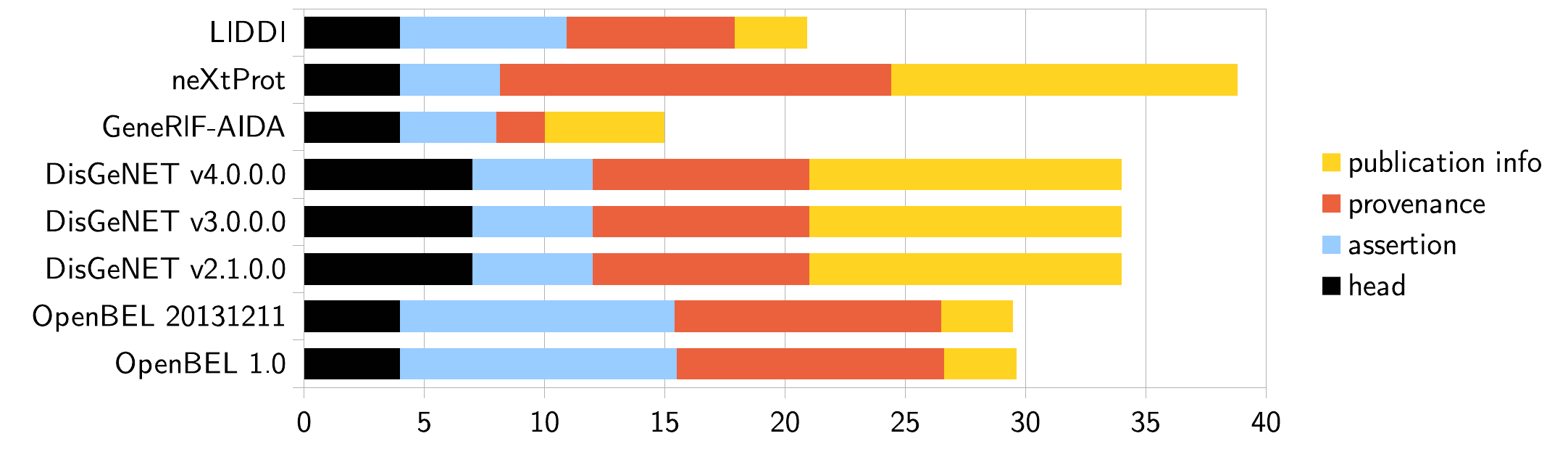}
\caption{Average triple counts of existing nanopublication datasets.}
\label{fig:nanopubinspection}
\end{center}
\end{figure}

\begin{table}[tb]
\begin{center}
\caption{Characteristics of existing nanopublication datasets.}
\label{tab:datasetinspection}
\begin{tabular}{r@{~~~}|@{~~~}r@{~~~}r@{~~~}r@{~~~}r@{~~~}r@{~~~}r}
 & & & triples & decontext- & \\
 & nanopub- & total & outside of & ualized & ratio \\
dataset & lications & triples & head ($t$) & triples ($d$) & $d/t$ \\
\hline
LIDDI & 98085 & 2051959 & 1659619 & 1364314 & 0.8221 \\
neXtProt & 4025981 & 156263513 & 140159589 & 76722914 & 0.5474 \\
GeneRIF-AIDA & 156026 & 2340390 & 1716286 & 733208 & 0.4272 \\
DisGeNET v4.0.0.0 & 1414902 & 48106668 & 38202354 & 5390141 & 0.1411 \\
DisGeNET v3.0.0.0 & 1018735 & 34636990 & 27505845 & 3908268 & 0.1421 \\
DisGeNET v2.1.0.0 & 940034 & 31961156 & 25380918 & 3667767 & 0.1445 \\
OpenBEL 20131211 & 74173 & 2186874 & 1890182 & 1308625 & 0.6923 \\
OpenBEL 1.0 & 50707 & 1502574 & 1299746 & 903066 & 0.6948 \\
\end{tabular}
\end{center}
\end{table}

Fine-grained and provenance-aware approaches like nanopublications, however, come at a cost. The internal structure of each nanopublication has to be defined, and the provenance and metadata has to be repeated even if it is virtually identical for a large number of them. This effect can be seen in Figure \ref{fig:nanopubinspection} for a number of existing dataset that use the nanopublication format: LIDDI \cite{banda2015provenance}, neXtProt \cite{chichester2015converting}, GeneRIF-AIDA \cite{kuhn2013broadening}, three versions of DisGeNET \cite{queralt2016publishing}, and two versions of a dataset extracted from OpenBEL\footnote{\url{https://github.com/tkuhn/bel2nanopub}}. We see that the nanopublication format implies a significant overhead in terms of number of triples. The main content of a nanopublication in the assertion graph account for just a minority of the total triples. While the provenance and publication info graphs provide additional context for the assertion triples, the head graph's sole purpose is to link to the other graphs and thereby to hold the nanopublication together.

While the provenance and publication information contents are by no means useless and therefore not purely an overhead, they tend to be quite repetitive. This is at least partly caused by the fact that most existing nanopublication datasets are extracted from ``non-nano'' datasets that do not capture granular metadata, and therefore no granular metadata is available for export. The overhead is in any case significant for existing datasets, as shown in Table \ref{tab:datasetinspection}. Even when disregarding the triples of the head graph, the numbers of triples is significantly larger than what we get if we ``decontextualize'' the triples to attach provenance and metadata only to the entire dataset and remove all duplicates. A decontexualized dataset, for example, would state that a given publication was the source of some entries in the dataset, but not refer to these exact entries, as enforced with nanopublications. We will use this method of \emph{decontextualization} also below for our analyses. DisGeNET is an extreme example here, with the number of decontextualized triples making up only 14\% of the number of nanopublication triples, caused by the repetition of triples across nanopublications.

This significant overhead that comes with the nanopublication technology might have been a hindrance in its further adoption. We show here, however, that nanopublications together with an approach to represent and construct incremental datasets and subsets thereof lead to a situation where the benefits of the fine-grained nanopublication structure offset the costs, even for the most extreme case of the DisGeNET dataset.

\section{Background}

Versioning and capturing the evolution of Linked Data has been a concern and research area for many years. While the early work focused on capturing the changes in ontologies \cite{volkel2005semversion,auer2006versioning}, later work included approaches to combine RDF versioning with web archiving \cite{van2010http}, long-term observation of the dynamics of Linked Data \cite{kafer2013observing}, and efficient archiving of dynamic Linked Data \cite{fernandez2015diacron}. There have also been a few approaches that deal with access and versions of \emph{subsets} of Linked Data resources \cite{schandl2010replication,silvello2015dlib}.

Providing version indicators for datasets is considered common best practice\footnote{see e.g. \cite{rauber2016tcdl} and \url{https://www.w3.org/TR/dwbp/\#dataVersioning}}, but version numbers cannot guarantee that data providers do not violate a dataset version's immutability. To provide such kinds of strong technical guarantees, approaches inspired by the Git versioning system have been proposed \cite{vandersande2013ldow,graube2014r43ples} that involve cryptographic hash values to enforce immutable versions. Similar approaches to reliable incremental Linked Data versioning have been developed by others \cite{meinhardt2015tailr,frommhold2016towards}, including applications to Big Data environments \cite{chard2016bigdata}. Outside of the Linked Data world, approaches for cryptographically strong data archiving have been proposed for decentralized systems like Bitcoin \cite{miller2014sp} and BitTorrent \cite{cohen2014xsede}.

In our own previous work, we showed how nanopublications with trusty URIs can make data publishing verifiable and reliable, without depending on a central server or trusted authority \cite{kuhn2015publishing}. In the same work, we also proposed a method to describe datasets as nanopublications themselves, thereby making references to entire sets of nanopublications verifiable through recursive hashing.

While a number of approaches exist on each of (1) Linked Data versioning, (2) cryptographically reliable dataset identifiers, and (3) references to subsets of larger datasets, and while these aspects are covered by the data citation recommendations of the Research Data Alliance \cite{rauber2016tcdl}, there are currently no concrete solutions that combine them all. In other words, existing approaches do not allow for cryptographically reliable references at high granularity in terms of both, time (i.e. versions) and space (i.e. subsets). We will present and evaluate such an approach below.

\section{Approach}

Our approach consists of the following three aspects: (1) We use the nanopublication concept to model datasets and their versions, (2) provide a method to create incremental datasets, and (3) connect these components to allow for flexible and reliable references to subsets of data resources.

\subsection{Incremental Datasets with Nanopublications}

\begin{figure}[tb]
\begin{center}
\includegraphics[width=\textwidth]{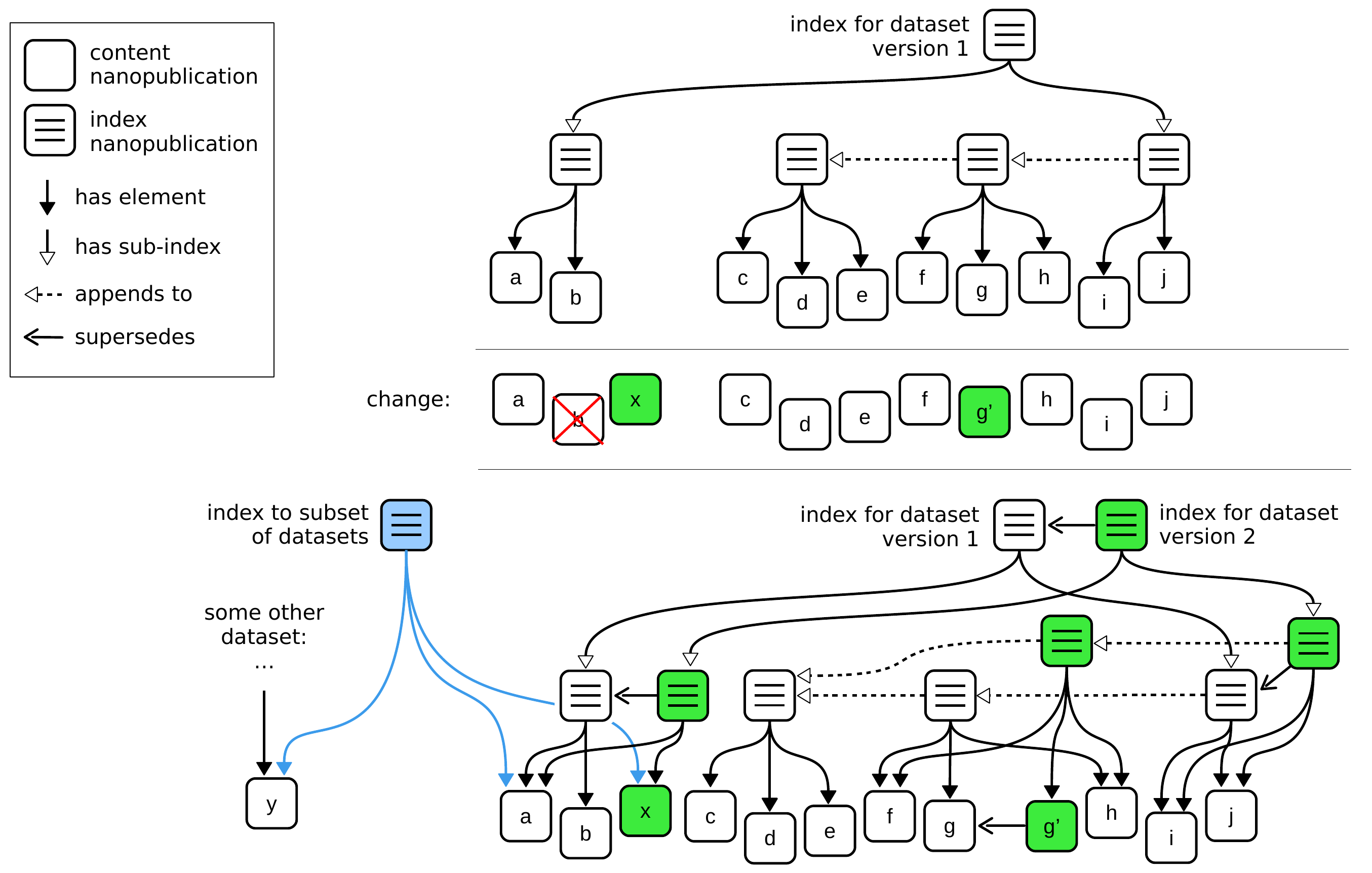}
\caption{Schematic depiction of a dataset specified with nanopublication indexes (top), the occurred content changes (middle), and their result as a new dataset version that reuses as much as possible. The blue index shows a subset definition.}
\label{fig:diagram}
\end{center}
\end{figure}

Figure \ref{fig:diagram} schematically depicts the gist of our approach. It is based on our previous proposal to define sets of nanopublications as nanopublications themselves \cite{kuhn2016peerjcs}. We call these set-defining nanopublications \emph{index nanopublications}, as they consist of direct and indirect links to the nanopublications they contain as elements. An index nanopublication can directly link to elements via links of the type \emph{has element} (these elements are marked with lowercase letters in Figure \ref{fig:diagram}), but can also point to subsets in the form of other indexes via links of the type \emph{has sub-index}. Sub-indexes can be used, for example, to partition a dataset into different parts each containing a particular type of data.
Finally, for nanopublication sets that are large but have no such partitioning, we need a method to ensure that all these index nanopublications remain small, as this is a core feature of the nanopublication concept. For that reason, we introduce relations of the type \emph{appends to} that allows for more nanopublications being added in a new index, once an index is full. The size limit of a nanopublication index is set to 1000 entries (either elements or sub-indexes). All these links are established via the trusty URIs of the referred nanopublication, and thereby the whole reference tree can be cryptographically verified from just the URI of the top index nanopublication \cite{kuhn2015tkde}. We will come back below to the issue of how to retrieve such sets of nanopublications.

Because of its granularity, this approach provides excellent opportunities to reuse parts of a dataset for a new version in an incremental manner. In general, there are three kinds of changes that can happen: A nanopublication can be removed from a dataset (such as $b$ in Figure \ref{fig:diagram}); a nanopublication can be added ($x$); and a nanopublication can be changed and replaced by a new version ($g$ being replaced by $g'$). All remaining nanopublications remain unchanged and can thereby be reused, i.e. linked from an index nanopublication belonging to the new version of the dataset.
Moreover, we might also be able to reuse some of the nanopublication indexes, namely the ones representing subsets that didn't change.
For both, content and index nanopublications, we can furthermore establish \emph{supersedes} links to the respective previous versions, to allow users to navigate back in time through the version history.

It is important to note that the previous version remains untouched: None of the existing nanopublications are changed (trusty URIs in fact enforce this) and by starting from the URI of the previous version and follow its links, the existence of the new version is not even noticed. Turning this property around implies that defining sets of nanopublications in this way does not require any control over the contained elements. Everybody can define after the fact (i.e. after the release of a dataset) arbitrary subsets by creating the appropriate index nanopublications. These subsets are maximally flexible in the sense that they can reuse any possible subset, be augmented with new nanopublications, and even combine subsets of different datasets, as illustrated by the blue index nanopublication in Figure \ref{fig:diagram}.
In such a case, one has to publish the new index nanopublications to be able to publicly refer to the specified subset, but no part of the content needs to be republished, and its original state is easy to verify.

We base our implementation and evaluation on the specific technologies and formats underlying Linked Data and nanopublications, but our general approach is portable to any type of knowledge representation with declarative monotonic semantics, which by their nature allow for subdividing representations into small independent pieces.

\subsection{From Snapshots to Incremental Datasets}

To actually generate an incremental dataset for a nanopublication-based resource, one has to ideally record all changes when they occur and build the proper index structure accordingly. However, such a direct construction is often non-trivial to integrate in existing data production pipelines, which is why first producing a full new snapshot and then calculating an incremental update is often more practical, in particular for smaller datasets. We therefore present such an approach here and apply it in the evaluations described below.

To calculate incremental updates of nanopublications, we apply the two concepts of \emph{fingerprints} and \emph{topics}. These two concepts establish identity relations that are weaker than the one that is enforced by trusty URIs. With trusty URIs, any tiny change in a nanopublication, such as a new timestamp, leads to a new URI and therefore to a new nanopublication. In contrast to trusty URIs, neither fingerprints nor topics are visible to the users of the dataset, but are merely a method to calculate incremental updates from dataset snapshots.

Fingerprints --- like trusty URIs --- correspond to a cryptographic hash value that is based on the RDF content of nanopublication, but consider only a subset of the triples and may apply preprocessing and normalization. In the simplest case, a fingerprint ignores the content of the timestamp found in the publication info graph. Other variants are possible, such as ignoring the entire publication info graph, and this can be configured for a given dataset and the intended use of its incremental versioning. The purpose of these fingerprints is to decide whether a new nanopublication (i.e. a nanopublication that would get a new trusty URI) is ``new enough'' to warrant an update, or whether a nanopublication from the previous version of the dataset can be reused.

Topics are similar to fingerprints, but normally correspond to a URI instead of a hash. A new nanopublication with an existing topic \emph{is} included in the new dataset version, but the new nanopublication will be marked as an update of the old. The addition of \emph{supersedes}-links as shown in Figure \ref{fig:diagram} thereby provides users a access to the version history on the level of individual nanopublications. By default, the topic is calculated to be the URI that has the highest occurrence in the subject position of the assertion triples, but this can be configured to match the characteristics of a given dataset.

It is worth noting that the matching of fingerprints and topics comes at a cost, in particular the cost of keeping a mapping table during the process. For large datasets, it can therefore pay off to record changes as they happen, which eliminates the need to reconstruct changes with fingerprints or topics.

\subsection{Granular and Reliable Retrieval}

So far we have only described our approach from a conceptual level assuming a reliable method to follow links. The most straight-forward approach to actually do this is the ``follow your nose'' principle\footnote{\url{https://www.w3.org/wiki/FollowYourNose}}
of URI dereferencing, which however is in general not reliable and can be very slow, depending on web servers a user has no control over. This problem is particularly grave for large datasets and those spanning multiple web domains. We also need to provide convenient methods for users to make their own subset definitions publicly available.

We address these problems by applying and using the decentralized server network that we demonstrated in previous work, based on nanopublications and trusty URIs \cite{kuhn2016peerjcs}. With this network, we do not have to assume that URIs are efficiently resolvable, but we can instead rely on the redundancy of the network and the verifiability of trusty URIs.
This nanopublication network has grown in the last months and years, consisting now of 15 server instances on 10 distinct physical servers in 8 countries.\footnote{\url{http://purl.org/nanopub/monitor}}
Our approach relies on this server network to let data producers publish incremental datasets, and to allow researchers to publish index nanopublications to precisely specify the subsets of existing Linked Data resources they are using for their analyses.

\section{Implementation and Methods}

We implemented our approach in a command line tool, and evaluated it with two studies. We performed a technical study covering the publishing aspect to find out about the overall data volume for changing datasets with our approach and to compare it to idealized alternative approaches of decontextualized triples. We then performed a second study to investigate how our approach performs on typical subsets of datasets that are used in scientific studies.

\subsection{Nanopublication Operation Tool: \emph{npop}}

Based on our existing \emph{nanopub-java} library\footnote{\url{https://github.com/Nanopublication/nanopub-java}} \cite{kuhn2015nanopub}, we implemented a command line tool that we call \emph{npop} (standing for \emph{n}ano\emph{p}ublication \emph{op}erations). The following commands are relevant to the work presented here:
\begin{itemize}
\item \texttt{count} can be used to count nanopublications and their triples from a file or stream. It is therefore like a \texttt{wc} command for nanopublications.
\item \texttt{filter} reads nanopublications from a file or stream and filters them by given URIs or literals. It is therefore like a \texttt{grep} command for nanopublications.
\item \texttt{extract} retrieves triples from the different nanopublication graphs.
\item \texttt{reuse} takes a dataset snapshot and its previous version, and generates an incremental update from it. Nanopublications from the previous version with a matching fingerprint are reused, and for those with a matching topic (but not a matching fingerprint) a \emph{supersedes}-link is introduced.
\item \texttt{ireuse} does the same as \texttt{reuse} but for index nanopublications.
\item \texttt{fingerprint} calculates the fingerprints for nanopublications following a specified configuration.
\item \texttt{topic} calculates the topics according to a specified configuration.
\item \texttt{decontext} produces decontextualized triples for given nanopublications, for comparative studies such as the ones presented in this paper.
\end{itemize}
These commands, together with the commands from the underlying \emph{nanopub-java} library (such as \texttt{get} to retrieve nanopublications and \texttt{publish} to upload them to the network), allowed us to perform the studies to be described below, and they are available for other data producers to apply to their own datasets.

\subsection{Evaluation on Data Publishing}

The first evaluation was performed on WikiPathways, a community-curated open database of
biological pathways~\cite{Kutmon2016}, with the aim to find out whether our approach is beneficial on the data producer side. Recently, the RDF export of the WikiPathways database was established~\cite{Waagmeester2017}, making the content of the database much easier to integrate. This RDF export contains information from the original WikiPathways and Reactome pathways~\cite{Fabregat2016,Bohler2016}. Using a number of SPARQL CONSTRUCT queries, three types of nanopublications are generated:\footnote{\url{https://github.com/wikipathways/nanopublications}} interactions, complex participation, and pathway participation. Importantly, only nanopublications are generated for statements if the fact is supported by a publication, marked with a PubMed database identifier. Overall, the dataset currently consists of a bit over 10\,000 nanopublications.

For this evaluation, we retroactively generated nanopublication snapshots from old data dumps, corresponding to 11 monthly builds between June 2016 to May 2017 (January 2017 is missing). For these we built an incremental dataset using the \emph{npop} tool. We can then compare the size of the resulting cumulative dataset, growing over 11 months, with the size of the nanopublication snapshots as well as decontextualized versions thereof, to evaluate whether incremental versioning can indeed offset the increased space needs of nanopublications.

This is not a very fair comparison, of course, because nanopublications come with valuable context-dependent information on the one hand and because incremental versioning could just as well be applied to decontextualized data on the other. We will keep the first point in mind when interpreting the results, and to address the second point we calculate an incremental version for the decontextualized case too.
Three general approaches exist for versioning of arbitrary RDF data \cite{tzitzikas2008storage,fernandez2015diacron}: independent copies, change-based approach, and timestamp-based approach. Independent copies correspond to what we called dataset snapshots, i.e. non-incremental versions. The change-based approach keeps separate lists of added and removed triples for each version after the first, whereas the timestamp-based approach keeps all triples in the same collection but attaches timestamps of their addition or removal. While the latter two have different advantages and shortcomings, they lead to the same overall triple count (if we require a triple to be duplicated to acquire more than one timestamp).
As a further point of comparison for our study, we therefore use this overall triple count for an incremental decontextualized dataset according to the change-based or timestamp-based approach.

\subsection{Evaluation on Data Analyses}

With the second evaluation we wanted to find out whether our approach is beneficial on the consumer end.
It was performed on DisGeNET \cite{pinero2016disgenet}, one of the most comprehensive databases on human diseases and their genes that is publicly available. DisGeNET is available in RDF \cite{queralt2016disgenet} and nanopublication \cite{queralt2016publishing} formats. There are currently three releases of the DisGeNET nanopublication dataset (version 2.1 with 940\,034 nanopublications, version 3.0 with 1\,018\,735 nanopublications, and version 4.0 with 1\,414\,902 nanopublications), which correspond to three most recent releases of the database. The releases differ mainly in data content due to the incremental update of the database, the incorporation of new data sources for the gene-disease associations, and the incorporation of new data attributes.

To find out about the use of this dataset by researchers, we looked at the publications that cited one of the DisGeNET papers during 2017 (31 publications as of 5 May 2017). We were interested in studies that included the DisGeNET dataset or subsets thereof in their analyses, but closer inspection revealed that six of these publications did not actually use the data (but only mentioned DisGeNET as related work) and another five of them used the data but did not include them in any analyses (e.g. describing a tool that imported the data).
For the remaining 20 publications, we manually determined whether the authors used the whole dataset or specific subsets. If the study used a specific subset, we looked for information about how this selection was performed (e.g. based on a particular disease or a family of genes, or using a pre-defined value of some of the DisGeNET data attributes as such as the DisGeNET score, among others).
Finally, we matched these subsets to the corresponding subsets of our incremental nanopublication-based dataset to find out what set of nanopublications they \emph{would have} used if they had followed our proposed approach.

From this empirical collection of used subsets, we can then investigate the typical size of such database subsets used for scientific analyses. We can also compare the size of these subsets to the decontextualized version of DisGeNET to find out whether the overhead of nanopublications is actually still an overhead once we look at specific subsets. We can reliably refer to such subsets with nanopublications, but we have to refer to (and therefore handle) the entire dataset for data based on regular (decontextualized) triples.

Finally, to measure the practicality of retrieving subsets from the server network, we also measure the time it takes to do so for a typical subset. To put that into perspective, we also measure the time needed to download the entire dataset from the \texttt{disgenet.org} website.

\section{Results}

\begin{table}[tb]
\begin{center}
\caption{Overview of the incremental dataset generated for WikiPathways.}
\label{tab:wpmonthly}
\begin{tabular}{l@{~}|@{~}r@{~}|@{~}rr@{~}rr@{~}|@{~}rr@{~}rr@{~}}
 & nanopub- & & & & & & & \\
version & lications & \multicolumn{2}{l}{reused} & \multicolumn{2}{l@{~}|@{~}}{new} & \multicolumn{2}{l}{update} & \multicolumn{2}{l}{addition} \\
\hline
20160610 & 9018 & 0 & (0.0\%) & 9018 & (100.0\%) & 0 & (0.0\%) & 9018 & (100.0\%) \\
20160710 & 10173 & 1405 & (13.8\%) & 8768 & (86.2\%) & 3 & (0.0\%) & 8765 & (100.0\%) \\
20160810 & 10123 & 3836 & (37.9\%) & 6287 & (62.1\%) & 0 & (0.0\%) & 6287 & (100.0\%) \\
20160910 & 10124 & 9838 & (97.2\%) & 286 & (2.8\%) & 0 & (0.0\%) & 286 & (100.0\%) \\
20161010 & 10127 & 9620 & (95.0\%) & 507 & (5.0\%) & 16 & (3.2\%) & 491 & (96.8\%) \\
20161110 & 13958 & 10041 & (71.9\%) & 3917 & (28.1\%) & 18 & (0.5\%) & 3899 & (99.5\%) \\
20161210 & 13975 & 13794 & (98.7\%) & 181 & (1.3\%) & 152 & (84.0\%) & 29 & (16.0\%) \\
20170210 & 14323 & 13743 & (96.0\%) & 580 & (4.0\%) & 176 & (30.3\%) & 404 & (69.7\%) \\
20170310 & 14319 & 13938 & (97.3\%) & 381 & (2.7\%) & 230 & (60.4\%) & 151 & (39.6\%) \\
20170410 & 14323 & 13972 & (97.5\%) & 351 & (2.5\%) & 317 & (90.3\%) & 34 & (9.7\%) \\
20170510 & 14323 & 13980 & (97.6\%) & 343 & (2.4\%) & 340 & (99.1\%) & 3 & (0.9\%) \\

\end{tabular}
\end{center}
\end{table}

Table \ref{tab:wpmonthly} gives an overview of the structure of the incremental dataset for WikiPathways, showing the number of nanopublications for each release, the number of reused nanopublications from the previous version (by fingerprint matching), and the number of new nanopublications. The right-hand side of the table shows how many of the new ones were updates of nanopublications from the previous version (by topic matching).
We see that the datasets underwent fundamental changes in the first two months, with a majority of nanopublications being replaced. Afterwards, the changes are much less drastic, in the sense that the majority of nanopublications are reused and often a majority of the new ones can be linked to previous nanopublications of the same topic.

\begin{figure}[tb]
\begin{center}
\includegraphics[width=0.95\textwidth]{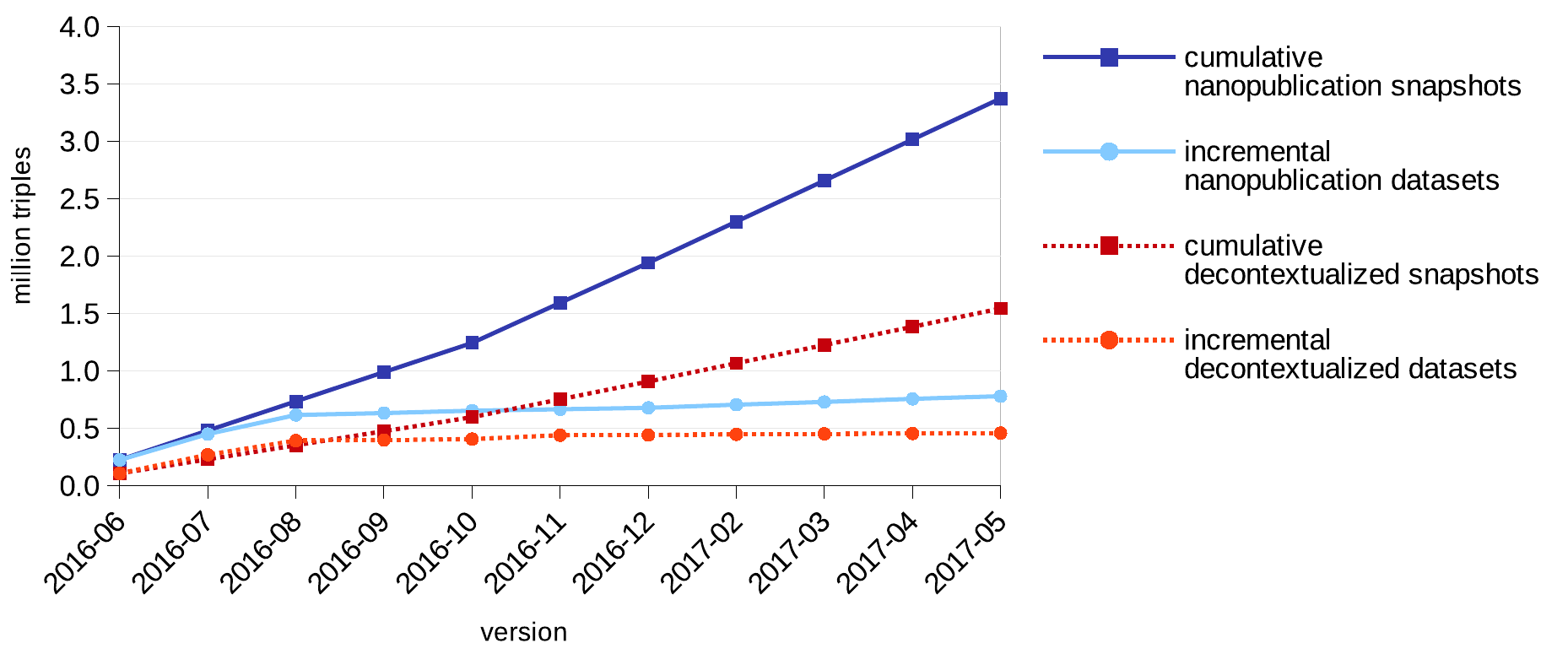}
\caption{Overall size of the evolving WikiPathways version history.}
\label{fig:wpresults}
\end{center}
\end{figure}

Figure \ref{fig:wpresults} shows the gains from the incremental approach to nanopublication-based versioning (light blue line). After the first two tumultuous months, the gain in number of triples to the cumulative nanopublication-based snapshots (dark blue line) quickly widens. In the end, we only need 23\% (0.78M/3.38M) of the triples to express the same version history. Comparing the two to our main reference point --- cumulative snapshots of decontextualized triples (dark red line) --- we see that the overhead of the nanopublication snapshots is in the end 54\% (1 -- 1.55M/3.38M), meaning that we could drop 54\% of the triples if we weren't interested in the fine-grained context. With the incremental nanopublication datasets, however, this overhead turns into a ``negative overhead'' of 98\% (1 -- 1.55M/0.78M), meaning that we needed 98\% more triples if we were to switch to decontextualized snapshots. We see that the overhead of nanopublications has indeed turned into a gain.

As we noted above, this comparison is not perfectly fair on either side. Still keeping in mind that decontextualized triples carry less information, we can compare our incremental nanopublication-based approach to what could be ideally achieved with a change-based or timestamp-based approach on decontextualized triples (light red line). The overhead of our approach to this idealized setting is 41\% (1 -- 0.46M/0.78M). The fact that this is again an actual overhead is not surprising, as it is always possible to handle \emph{less} information more efficiently. We will show below, however, that even this overhead is in fact turned into a gain when we look at the side of data consumers and the typical subsets they use.

\begin{table}[tb]
\begin{center}\footnotesize
\caption{DisGeNET subsets used and reported in papers, sorted by ascending size.}
\label{tab:disgenetsubsets}
\begin{tabular}{l@{~~~}|@{~~~}r@{~~~}r@{~~~}r@{~~~}r@{~}c}
 & nanopub- & & rel. size & rel. size to \\
 & lication & triple & to full & decontext. \\
DOI of paper & count & count & dataset & version \\
\hline
10.21873/cgp.20028 & 14 & 476 & 0.00001 & \cellcolor{green!25} 0.00009 \\
10.3892/ijmm.2017.2853 & 482 & 16388 & 0.00034 & \cellcolor{green!25} 0.00304 \\
10.1007/s12539-017-0213-z & 533 & 18122 & 0.00038 & \cellcolor{green!25} 0.00336 \\
10.1038/srep46760  & 782 & 26588 & 0.00055 & \cellcolor{green!25} 0.00493 \\
10.1016/j.preteyeres.2017.02.001 & 1711 & 58174 & 0.00121 & \cellcolor{green!25} 0.01079 \\
10.1101/gr.210740.116  & 2014 & 68476 & 0.00142 & \cellcolor{green!25} 0.01270 \\
10.1186/s12920-017-0259-0 & 2158 & 73372 & 0.00153 & \cellcolor{green!25} 0.01361 \\
10.1016/j.jprot.2017.03.015 & 4859 & 165206 & 0.00343 & \cellcolor{green!25} 0.03065 \\
10.1016/j.neuron.2017.01.033 & 18098 & 615332 & 0.01279 & \cellcolor{green!25} 0.11416 & * \\
10.1021/acs.jcim.6b00725 & 21336 & 725424 & 0.01508 & \cellcolor{green!25} 0.13458 \\
10.1101/119099 & 31105 & 1057570 & 0.02198 & \cellcolor{green!25} 0.19620 \\
10.1002/jcb.25799 & 61198 & 2080732 & 0.04325 & \cellcolor{green!25} 0.38603 \\
10.3390/ncrna3020020 & 78742 & 2677228 & 0.05565 & \cellcolor{green!25} 0.49669 \\
10.1007/978-1-4939-6843-5\_13 & 83771 & 2848214 & 0.05921 & \cellcolor{green!25} 0.52841 \\
10.1038/srep43632 & 101297 & 3444098 & 0.07159 & \cellcolor{green!25} 0.63896 \\
10.1016/j.dib.2017.04.001 & 196108 & 6667672 & 0.13860 & \cellcolor{red!25} 1.23701 \\
10.1186/s13148-017-0336-4 & 326472 & 11100048 & 0.23074 & \cellcolor{red!25} 2.05932 \\
10.1038/srep40154 & 1414902 & 48106668 & 1.00000 & \cellcolor{red!25} 8.92494 \\
10.1038/srep42638 & 1414902 & 48106668 & 1.00000 & \cellcolor{red!25} 8.92494 \\
10.1002/pmic.201700056 & 1414902 & 48106668 & 1.00000 & \cellcolor{red!25} 8.92494 \\
\hline
average: & 258769 & 8798156 & 0.18289 & 1.63227 \\
median: & 26221 & 891497 & 0.01853 & 0.16539 \\
\hline
average of proper subsets: & 54746 & 1861360 & 0.03869 & 0.34533 \\
median of proper subsets: & 18098 & 615332 & 0.01279 & 0.11416 \\
\end{tabular}
\end{center}
\end{table}

\begin{figure}[tb]
\begin{center}
\includegraphics[trim=0mm 5mm 10mm 20mm,width=0.9\textwidth]{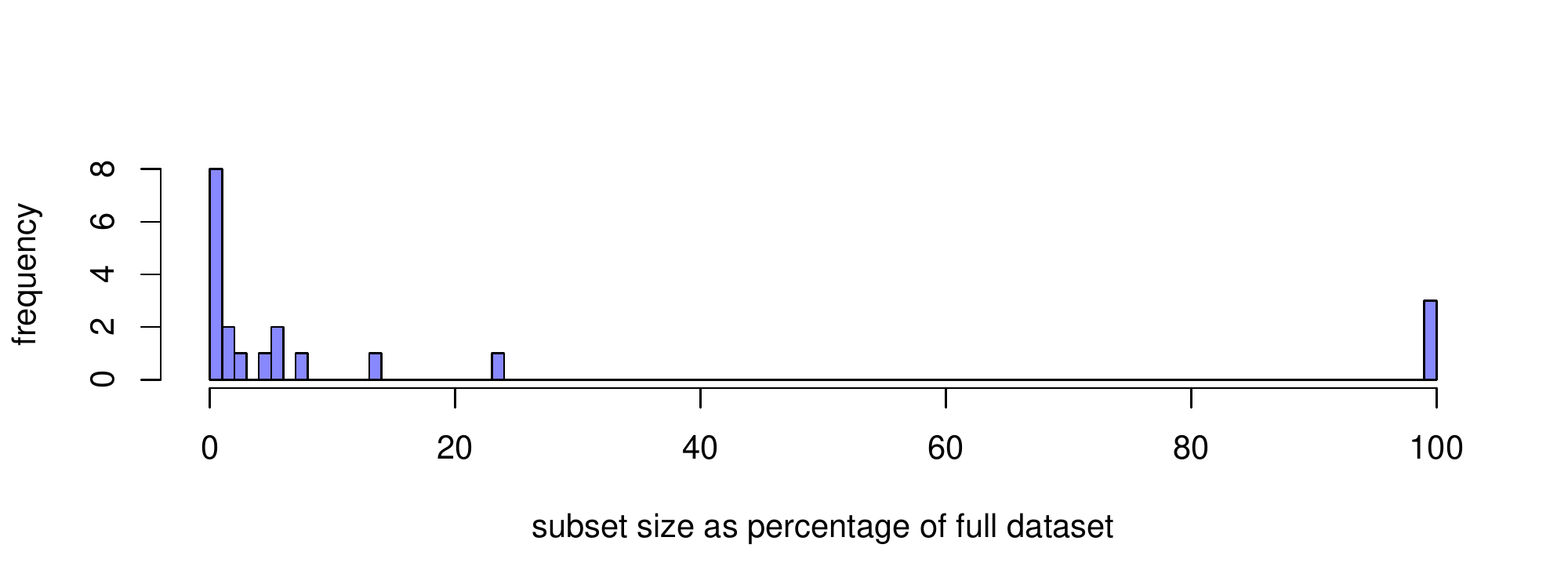}
\caption{Histogram of the subset sizes (in triples) in relation to the entire dataset.}
\label{fig:subsethist}
\end{center}
\end{figure}

Table \ref{tab:disgenetsubsets} shows the result of the second empirical study on the subsets of DisGeNET used and reported in scientific papers from 2017. Only three out of the 20 papers used the entire dataset. The distribution of the subset sizes is also shown in Figure \ref{fig:subsethist} as a histogram. The two peaks indicate that researchers tend to use a dataset either entirely or only a very small subset of it. For 40\% of the papers we studied (8 out of 20), less than 1\% of the dataset was used. The largest proper subset used consisted of just 23\% of the data.

We can again compare these numbers to the idealized setting without nanopublications where triples are decontextualized and where reliable identifiers only exist at the dataset level. In comparison to such a decontextualized snapshot, 15 out of the 20 studied subsets have a lower triple count (green). For a typical subset, the overhead of nanopublications in terms of number of triples is therefore again turned into a gain (in addition to the gains with respect to precision, verifiability and fine-grained provenance and metadata).
We should also remember that DisGeNET is an extreme case in terms of triple overhead.

\begin{figure}[tb]
\begin{center}
\includegraphics[width=0.95\textwidth]{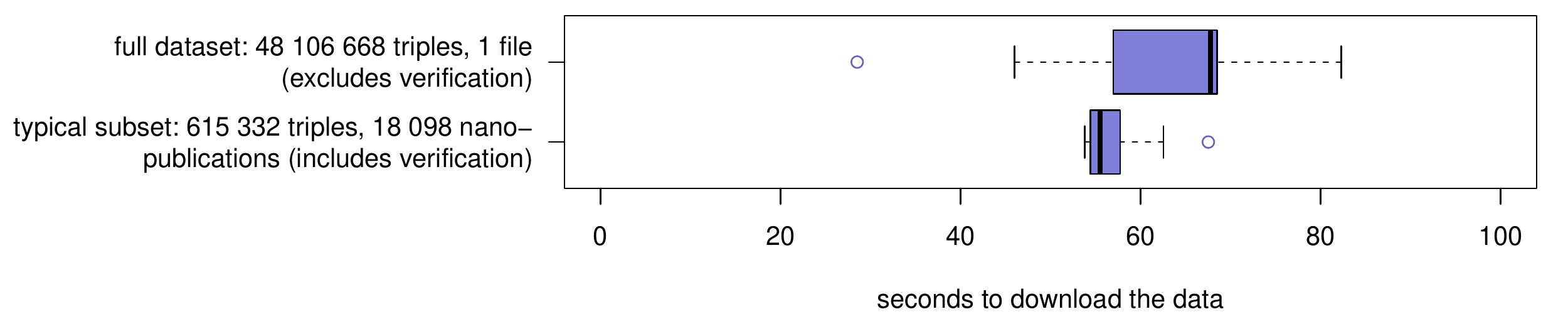}
\caption{Download times for the full DisGeNET dataset (v4.0.0.0) and a typical subset (marked with * in Table \ref{tab:disgenetsubsets}; $n$=10 in both cases; whiskers show +/-- 1.5 IQR).}
\label{fig:timeeval}
\end{center}
\end{figure}

Finally, Figure \ref{fig:timeeval} shows the results for the retrieval times of a typical subset (the subset with the median size value of the proper subsets, marked with * in Table \ref{tab:disgenetsubsets}). We see that the retrieval via the server network takes about the same time as downloading the whole dataset from \texttt{disgenet.org} (both roughly around 60 seconds). Instead of just downloading a single file, the subset retrieval consists of requesting 18\,098 individual nanopublications and verifying their content against their trusty URIs. Despite the resulting lower throughput in terms of triples per second, we can efficiently retrieve the specific subset of data.

The code used for these studies and the resulting data can be found online.\footnote{see \url{https://doi.org/10.6084/m9.figshare.5230639} and \url{https://bitbucket.org/tkuhn/nanodiff-exp/}}

\section{Discussion and Conclusions}

Data providers and data consumers have to pay a price for granular and precise references to subsets of datasets, to make these references cryptographically strong, and to verify the integrity of retrieved data. We showed, however, that this price is offset by the benefits of incremental versioning and by being able to refer to exactly the needed subset of a given dataset, on top of the gains from cryptographically strong verifiability. Data providers should take into account the gain in storage overhead and the benefits of reproducibility and verifiability --- and thus better FAIR publishing --- of evolving datasets that our incremental nanopublication approach provides. Also, it allows data publishers to reliably check and record how their data evolves from version to version.

To come back to the examples of dataset references, we can now refer to our datasets in papers with references that include the trusty URI of the nanopublication index of the appropriate version and subset, such as the incremental DisGeNET datasets \cite{nanopubindex2017disgenet2,nanopubindex2017disgenet3,nanopubindex2017disgenet4} and the incremental WikiPathways dataset \cite{nanopubindex20170510wikipathways} we cite in this paper. For integration in the code to perform computational analyses, we can now use the \texttt{np} command provided by the \emph{nanopub-java} library to reliably download a precisely specified set of nanopublications:
\begin{quote}\scriptsize
\texttt{np get -c -o data.trig http://purl.org/np/RAxMyDRaM8RmKGNiEe7dQPRUTuz616iI-N2T-H3MPYmXk}\\
\texttt{\# Run analysis here}
\end{quote}
We now get cryptographic guarantees on the retrieved content, and we can rely on an entire network of nanopublication servers and therefore do not depend on the uptime of individual servers.

As future work, we will keep providing incremental updates for the nanopublication datasets we presented here. We will also investigate how we can reduce the overhead present in DisGeNET nanopublications for future releases. The most obvious improvement is the reduction of the number of head triples from 7 to the mandatory minimum of 4. This alone will reduce the overall triple count by 9\%. Further improvements can probably be achieved --- without substantial negative side effects --- by reducing the redundancy in the provenance and publication info graphs, and possibly also in the assertion graph.

To conclude, we demonstrated how our approach can contribute to the verifiability and granular accessibility of scientific Linked Data resources. As such, we think that it can put many other Linked Data solutions that require precise and reliable data publishing and consumption onto a solid technical basis.

\section*{Acknowledgments}

We would like to thank Javier D. Fern\'andez for valuable input and discussions on RDF versioning.
L.I. Furlong and E. Centeno received support from ISCIII-FEDER (PI13/00082, CP10/00524, CPII16/00026), the EU H2020 Programme 2014-2020 under grant agreements no. 634143 (MedBioinformatics) and no. 676559 (Elixir-Excelerate). 

\bibliographystyle{abbrv}
\bibliography{references}

\end{document}